\documentclass[twocolumn,showpacs,preprintnumbers,amsmath,amssymb]{revtex4-1}
\usepackage{graphicx}% Include figure files
\usepackage{dcolumn}% Align table columns on decimal point
\usepackage{bm}% bold math
\usepackage{color}
\def\vc#1{\mbox{\boldmath $#1$}}

\begin{document}

\preprint{}

\title{The nuclear electric dipole moment in the cluster model with a triton: $^7$Li and $^{11}$B}% Force line breaks with \\

\author{Nodoka~Yamanaka$^{1,2}$}
  \email{ynodoka@yukawa.kyoto-u.ac.jp}
\author{Taiichi Yamada$^3$}
\author{Yasuro Funaki$^3$}
  \affiliation{$^1$Yukawa Institute for Theoretical Physics, Kyoto University, %\\
  Kitashirakawa-Oiwake, Kyoto 606-8502, Japan}
  \affiliation{$^2$IPNO, Universit\'{e} Paris-Sud, CNRS/IN2P3, %\\
 F-91406, Orsay, France}
%  \email{ynodoka@yukawa.kyoto-u.ac.jp}
   \affiliation{$^3$Laboratory of Physics, Kanto Gakuin University,%\\
Yokohama 236-8501, Japan}

\date{\today}% It is always \today, today,

\begin{abstract}
We calculate the electric dipole moment (EDM) of the nuclei $^7$Li and $^{11}$B in the cluster model with $\alpha$ ($^4$He) and triton ($^3$H) clusters as degrees of freedom.
The $^7$Li and $^{11}$B nuclei are treated in the two- and three-body problem, respectively, using the Gaussian expansion method, assuming the one-meson exchange P, CP-odd nuclear forces.
We find that $^7$Li and $^{11}$B have larger sensitivity to the CP violation than the deuteron.
It is also suggested that the EDMs of $^7$Li and $^{11}$B, together with those of $^6$Li, $^9$Be and the $1/2^+_1$ excited state of $^{13}$C, obey an approximate counting rule accounting for the EDM of the cluster and the $\alpha -N $ polarization.
We show their sensitivity on the hadronic level CP violation in terms of the chiral effective field theory, and discuss their role in probing new physics beyond the standard model.
\end{abstract}

\maketitle

\section{Introduction}

The common understanding of the matter abundance of the current Universe is that it was generated in its very early stage.
Theoretically, the excess of matter over antimatter is realized by fulfilling Sakharov's criteria \cite{sakharov}.
In the standard model (SM), however, the CP violation, which is one of the required conditions, is very insufficient \cite{shaposhnikov1,shaposhnikov2,farrar,huet}.
This fact is motivating particle physicists to search for CP violation beyond the SM.

One of the most promising experimental observables to search for CP violation is the {\it electric dipole moment} (EDM).
It has been extensively studied in the past \cite{hereview,Bernreuther,barrreview,khriplovichbook,ginges,pospelovreview,fukuyama,engelreview,yamanakabook,hewett,devriesreview,yamanakanuclearedmreview,atomicedmreview,chuppreview,safronova,orzel}.
Notable advantages are its accurate measurability, the negligible SM contribution \cite{smtheta,mannel,smneutronedmmckellar,czarnecki,seng,yamanakasmcpvnn,yamanakasmdeuteronedm,Lee:2018flm} from the CP violation of the Cabibbo-Kobayashi-Maskawa (CKM) matrix \cite{ckm}, and the versatility of the choice of systems, illustrated by the results obtained for the neutron \cite{baker}, atoms \cite{regan,graner,parker,Allmendinger:2019jrk}, muon \cite{muong-2}, and the electron in polar molecules \cite{Hudson:2011zz,Cairncross:2017fip,Andreev:2018ayy}.

Many new ideas for the measurement of the EDM were also proposed.
They include that of the strange and charmed baryons using bent crystals to extract the EDM of strange and charm quarks \cite{botella,Baryshevsky:2018dqq,Baryshevsky:2019vou,Baryshevsky:2019rlk} (a recent investigation shows that the EDM of heavy quarks can strongly be constrained by the neutron EDM bound through renormalization group analysis \cite{Gisbert:2019ftm}), precision analysis in collider experiments for the EDM of $\tau$ leptons \cite{xinchen,koksal,Koksal:2018xyi,Gutierrez-Rodriguez:2019umw} and top quarks \cite{Hernandez-Ruiz:2018oxq,Koksal:2019cjn}, the improvement of the sensitivity to electron EDM using polar molecules and an inert gas matrix \cite{Malika:2019jhn,inertgasmatrix}, the use of oscillating electric field to measure the nuclear EDM \cite{Flambaum:2018wkp,Dzuba:2018wei,Tan:2018ihu}, etc.
In this work, we focus on the nuclear EDM which may be measured by using storage rings \cite{hewett,yamanakanuclearedmreview,storage1,storage2,storage3,storage4,storage5,storage6}.
Although the idea is not new, this approach is currently being developed, and the experimental preparation is on-going \cite{Anastassopoulos,jedi,jedi2,bnl}.

Let us list the advantages of the nuclear EDM.
First, the sensitivity of the experiment is very high, with the prospect of going beyond $O(10^{-29})\, e$ cm \cite{bnl}.
The use of storage rings is not only important in improving the accuracy, but also because it allows the measurement of the EDM of charged particles.
In contrast to atoms or molecules, the measurement of the EDM of bare nuclei does not suffer from Schiff's screening \cite{schiff}.
The sensitivity is so high that the nuclear EDM can also be used to test the Lorentz violation \cite{Araujo:2015zsa,Araujo:2016hsn,araujo,Araujo:2019txk}.
Another potentially important aspect is that the axion dark matter \cite{Duffy:2009ig} may also be studied.
It was recently claimed that the EDM of atomic nuclei is not completely screened in an oscillating electric field \cite{Flambaum:2018wkp,Dzuba:2018wei,Tan:2018ihu}, and in the same logic, the oscillation of the nuclear EDM induced by the interaction with the axion of the dark matter halo surrounding us might also be observable \cite{Graham:2011qk,Stadnik:2013raa,Abel:2017rtm,Flambaum:2019emh}.
All these topics motivate us to study the nuclear EDM in detail. 

There have recently been several notable progresses in the physics closely related to the nuclear EDM.
It has been pointed out that the measurement of the EDM in a storage ring may be affected by the systematics due to the general relativity \cite{Kobach:2016kvn,Silenko:2006er,Obukhov:2016vvk,Obukhov:2016yvw,Orlov,Laszlo:2018llb,Laszlo:2019cyu}.
This effect can be removed by measuring oppositely circulating particles in the ring.
At the hadronic level, there are many investigations in lattice QCD trying to evaluate unknown hadron matrix elements which are representing the most important source of theoretical uncertainty in the calculation of the EDM \cite{dragos,detmold}.
It is now possible to calculate several of them at the physical pion mass, such as the nucleon scalar and tensor charges \cite{chiqcdsigmaterm,rqcdsigmaterm,bmwsigmaterm,etmsigmaterm,Gubler:2018ctz,green,rqcdisovector,pndmeisovector,chiqcdisovector,etmsigmaterm,pndmetensor,egerer,Hasan:2019noy,Harris:2019bih}.
The calculation of the matrix elements of other operators such as the quark chromo-EDM or the $\theta$-term is still challenging \cite{ohki}, but attempts to obtain accurate values by using chiral effective field theory \cite{devries3} combined with high energy QCD experimental data are now starting \cite{seng2}.

The nuclear EDM has widely been investigated in nuclear physics.
Initial studies pointed to a large enhancement of CP violation for the EDM of relatively heavy nuclei, which can be described with the shell model \cite{sushkov1,sushkov2,sushkov3}.
More recent calculations based on a more sophisticated shell model accounting for the configuration mixing are however giving much smaller sensitivity, which is explained by the destructive interference of different angular momentum configurations \cite{yoshinaga1,yoshinaga2}.
From these inspections, light nuclei become more interesting in the context of the nuclear EDM measurement.

The EDM of the deuteron \cite{avishai1,korkin,pospelovdeuteron,liu,devries,dedmtheta} and $^3$He \cite{avishai2,avishai3,stetcu,chiral3nucleon,song} were studied in many previous works.
For these light nuclei, {\it ab initio} calculations are possible, and recent results are almost consistent \cite{afnan,bsaisou,bsaisou2,mereghetti,yamanakanuclearedm}.
The investigations further continued for $p$-shell nuclei, where the cluster model was used \cite{yamanakanuclearedm,c13edm}.
The calculation of the many-nucleon wave function is problematic, since it becomes computationally too costly even for a small number of nucleons such as the $^6$Li nucleus.
The cluster model can reduce the degree of freedom of the many-body system while keeping reasonable accuracy for low-lying energy levels.
It is also known that light nuclei have well developed cluster structures, so that the use of this framework matches the relevant physics \cite{clusterreview1,clusterreview2,clusterreview3,Hiyama:2009zz}.
The study of the EDM of $^6$Li and $^9$Be revealed the importance of the constructive interference between the EDM of subsystems and the polarization of the $\alpha-N$ system, which enhance the CP violation \cite{yamanakanuclearedm}.
On the other hand, the calculation of the EDM of $^{13}$C showed us that the nuclear EDM is suppressed due to the weak transition between low lying opposite parity states
\cite{c13edm}.
The results of these analyses tell us that the enhancement or the suppression depends strongly on the structure of the nuclei, and that we have to check them one by one.

In this work, we calculate the EDM of $^7$Li and $^{11}$B in the cluster model.
The $^7$Li \cite{nishioka,furutani} and $^{11}$B \cite{yamada11b} nuclei were calculated with $\alpha$-clusters and a triton ($^3$H) as degrees of freedom, and the reproduction of the low lying energy spectra was successful.
As hadronic CP violating processes generating the nuclear EDM, we consider the CP-odd one-pion exchange interaction \cite{Barton,pvcpvhamiltonian1,pvcpvhamiltonian2,pvcpvhamiltonian3} and the intrinsic nucleon EDM, which are the leading contribution in the chiral effective field theory \cite{bsaisou,bsaisou2,eft6dim}.
Since the one-pion exchange is a long distance process, we expect the cluster model, which reproduces well low lying energy levels, to accurately predict the nuclear EDM.
We use the Gaussian expansion method \cite{hiyama,Hiyama2012ptep} to solve the few-body problem. 

This paper is organized as follows.
In Sec. \ref{sec:nuclearedm}, we first define the nuclear EDM.
We then briefly introduce the cluster model and present our setup of the calculation in Sec. \ref{sec:setup}.
In Sec. \ref{sec:results}, we show and analyze our result by comparing with predictions of other nuclei.
In the final section, we summarize our discussion.

\section{The nuclear electric dipole moment\label{sec:nuclearedm}}

The first leading contribution to be considered in the calculation of the nuclear EDM is that of the intrinsic nucleon EDM.
It is given by 
\begin{eqnarray}
d_A^{\rm (Nedm)} 
&=&
\sum_{i}^A
d_i \langle \, \Psi_A\, |\, \sigma_{iz} \, |\, \Psi_A\, \rangle
\nonumber\\
&\equiv &
\langle \sigma_p \rangle_A \, d_p + \langle \sigma_n \rangle_A \, d_n
,
\label{eq:intrinsisnucleonedm}
\end{eqnarray}
where $|\, \Psi_A\, \rangle$ is the state vector of $^7$Li or $^{11}$B.
The mass number is denoted by $A$. 
The EDMs of the proton and the neutron, which are the input parameters, are denoted by $d_p$ and $d_n$, respectively.
The nuclear spin matrix elements $\langle \sigma_p \rangle_A$ and $\langle \sigma_n \rangle_A$ are the coefficients relating the nucleon EDM and the nuclear EDM.
These are one of the objects of our work, and they will be calculated in the cluster model.

Since we work in the cluster model with a triton, the nuclear spin matrix elements are given as a convolution of the spin matrix elements of the triton by the triton spin matrix elements in $^7$Li or $^{11}$B.
Regarding the spin matrix element of the triton, we will adopt the numerical value given in Ref. \cite{yamanakanuclearedm}.

We note that the intrinsic nucleon EDM contribution is not enhanced by the many-body effect, since nucleons in nuclei are nonrelativistic \cite{khriplovichbook,ginges,sandars1,sandars2,flambaumenhancement}.
The contribution from the interaction between the nucleon EDM and the internal electric field of the nucleus \cite{inoue} is also neglected.

The nuclear EDM is also generated by the CP-odd nuclear force which polarizes the total system.
We define this contribution as
\begin{eqnarray}
d_{A}^{\rm (pol)} 
&=&
\sum_{i=1}^{A} \frac{e}{2} 
\langle \, \tilde \Psi_A \, |\, (1+\tau_i^z ) \, {\cal R}_{iz} \, | \, \tilde \Psi_A \, \rangle
,
\end{eqnarray}
where $\frac{e}{2}(1+\tau_i^z )$ and ${\cal R}_{iz}$ are the charge operator and the third component of the coordinate of the $i$th nucleon, respectively.
Note that the EDM is calculated in the center-of-mass frame.
The nuclear state $|\, \tilde \Psi_A\, \rangle$ contains small opposite parity components due to the CP-odd nuclear force.

In the cluster model, the dipole operator is further simplified.
In our cluster model, the one of $^7$Li is given by
\begin{equation}
\sum_{i=1}^{7} \frac{e}{2} (1+\tau_i^z ) \, {\cal R}_{iz}
=
-\frac{2}{7} e ({\vc r}_1 -{\vc r}_2 )_z
,
\end{equation}
where ${\vc r}_1$ and ${\vc r}_2$ are the center-of-mass coordinates of the $^3$H and $\alpha$ clusters, respectively.
The origin is taken as the center of mass $3{\vc r}_1 + 4{\vc r}_2 =0$.

For the case of $^{11}$B, we consider the $\alpha -\alpha -^3$H three-body model.
Its dipole operator is then given by
\begin{equation}
\sum_{i=1}^{11} \frac{e}{2} (1+\tau_i^z ) \, {\cal R}_{iz}
=
-\frac{2}{11} e [ ({\vc r}_1 -{\vc r}_2)_z +({\vc r}_1 -{\vc r}_3)_z ]
,
\end{equation}
where ${\vc r}_1 -{\vc r}_2$ and ${\vc r}_1 -{\vc r}_3$ are the relative coordinates between $^3$H and the two $\alpha$ clusters.
Since the $\alpha$-clusters are bosons, the wave function of $^{11}$B is symmetric under the interchange ${\vc r}_2 \leftrightarrow {\vc r}_3$.

\section{Model setup\label{sec:setup}}

In this work, we use the Orthogonality Condition Model (OCM) 
which is the semi-microscopic approximation of the Resonating Group Method  \cite{ocm1,ocm2,horiuchi1,ocm3,horiuchi2}.
We first show the OCM Hamiltonian used in this work.
For the $^7$Li nucleus, it is given by
\begin{eqnarray}
\mathcal{H}_{^{7}{\rm Li}}
&=&
\sum_{i=1}^{2} T_i-T_{\rm cm}(^7{\rm Li})
+ V_{{\alpha}t}({\vc r}_1 -{\vc r}_2) 
\nonumber \\
&& 
+ V_{\rm Pauli} (^{7}{\rm Li})
+ V_{CP}({\vc r}_1 -{\vc r}_2)
,
\label{eq:7Lihamiltonian}
\end{eqnarray}
where $T_i-T_{\rm cm}$ stands for the kinetic energy operator for the {\it i}-th cluster from which the center-of-mass contribution has been subtracted, and $V_{{\alpha}t}({\vc r}_1 -{\vc r}_2)$ is the CP-even $\alpha-t$ interaction of Nishioka {\it et al.} \cite{nishioka,furutani}, which reproduces well the energy levels $3/2_1^- , 1/2_1^- , 7/2_1^-$, and $5/2_1^-$ of $^7$Li.
The forbidden states are projected out using the following Pauli-blocking operator $V_{\rm Pauli}(^{7}{\rm Li})$ \cite{Kukulin}:
\begin{eqnarray}
V_{\rm Pauli} (^{7}{\rm Li})
&=& 
\lim_{\lambda \rightarrow  \infty}\ {\lambda}\ \times
\nonumber\\
&& 
\sum_{2n+l < 3 }\sum_{i=2 }^3\ {| u_{nl} ({\vc r}_i -{\vc r}_1) \rangle}{\langle u_{nl}({\vc r}_i -{\vc r}_1) |}
,
\nonumber\\
\end{eqnarray}
which removes the $\alpha-t$ relative states $0S , 0P, 0D$, and $1S$.
In this work, we take $\lambda=10^{5}$~MeV.
The detail of the CP-odd nuclear force $V_{CP}$ will be given later.
For $^{11}$B, we have 
\begin{eqnarray}
\mathcal{H}_{^{11}{\rm B}}
&=&
\sum_{i=1}^{3} T_i-T_{\rm cm} (^{11}{\rm B})
+ \sum_{i=2}^{3} V_{{\alpha}t}({\vc r}_1 -{\vc r}_i) 
\nonumber \\
&& 
+ V_{{2\alpha}}(|{\vc r}_2 -{\vc r}_3|) 
+ V_{2\alpha t}({\vc r}_1 ,{\vc r}_2,{\vc r}_3) 
\nonumber\\
&& 
+ V_{\rm Pauli} (^{11}{\rm B})
+ \sum_{i=2}^{3} V_{CP}({\vc r}_1 -{\vc r}_i) 
,
\label{eq:11Bhamiltonian}
\end{eqnarray}
where $V_{2\alpha}$ is the effective $\alpha-\alpha$ potential constructed by folding the modified Hasegawa-Nagata interaction \cite{hasegawa} augmented by the Coulomb interaction.
This effective $\alpha-\alpha$ interaction reproduces well the $\alpha-\alpha$ scattering phase shifts as well as the energies of the $^8$Be ground-band states and of the Hoyle state ($0^+_2$ of $^{12}$C) \cite{yamada-schuck,funaki2}.
We also introduce a phenomenological three-body force $V_{2\alpha t}$ to fit the energies of $3/2_1^-$ and $1/2^-_1$ states of $^{11}$B \cite{yamada11b}.
The Pauli-blocking operator for $^{11}{\rm B}$ is given by
\begin{eqnarray}
V_{\rm Pauli} (^{11}{\rm B})
&=& 
\lim_{\lambda \rightarrow  \infty}\ {\lambda}\ \times
\nonumber\\
&&
\Biggl[
\sum_{2n+l < 4 , l={\rm even}}\hspace{-1.5em}
{| u_{nl} ({\vc r}_2 -{\vc r}_3) \rangle}{\langle u_{nl} ({\vc r}_2 -{\vc r}_3) |}
\nonumber\\
&&
+\sum_{2n+l < 3 }\sum_{i=2 }^3\ {| u_{nl} ({\vc r}_i -{\vc r}_1) \rangle}{\langle u_{nl}({\vc r}_i -{\vc r}_1) |}
\Biggr]
,
\nonumber\\
\end{eqnarray} 
which removes the $\alpha$-$\alpha$ relative states $0S,1S$, and $0D$, and the $\alpha-t$ relative states $0S , 0P, 0D$, and $1S$, with the same $\lambda$ as for $^7$Li.

\begin{figure}[htb]
\begin{center}
\includegraphics[width=8.5cm]{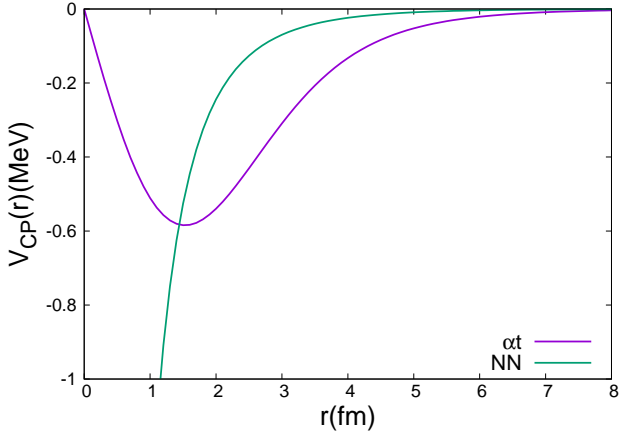}
\caption{\label{fig:comparison_vcp_folded_alpha-triton}
The CP-odd one-pion exchange $\alpha - t$ potential.
The CP-odd $N - N$ interaction is also shown for comparison.
The coupling constant $\bar G^{(1)}_\pi$ was factored out.
}
\end{center}
\end{figure}

As for the CP-odd nuclear force, we assume the one-pion exchange process \cite{korkin,liu,stetcu,song,bsaisou2,yamanakanuclearedm}.
The bare CP-odd nucleon-nucleon interaction is given by \cite{pvcpvhamiltonian1,pvcpvhamiltonian2,pvcpvhamiltonian3,liu}
\begin{eqnarray}
H_{P\hspace{-.5em}/\, T\hspace{-.5em}/\, }^\pi
& = &
\bigg\{ 
\bar{G}_{\pi}^{(0)}\,{\vc{\tau}}_{1}\cdot {\vc{\tau}}_{2}\, {\vc{\sigma}}_{-}
+\frac{1}{2} \bar{G}_{\pi}^{(1)}\,
( \tau_{+}^{z}\, {\vc{\sigma}}_{-} +\tau_{-}^{z}\,{\vc{\sigma}}_{+} )
\nonumber\\
&&\hspace{1em}
+\bar{G}_{\pi}^{(2)}\, (3\tau_{1}^{z}\tau_{2}^{z}- {\vc{\tau}}_{1}\cdot {\vc{\tau}}_{2})\,{\vc{\sigma}}_{-} 
\bigg\}
\cdot
\hat{ {\vc r}} \,
V_{CP}^{NN}(r)
,
\ \ \ 
\label{eq:CPVhamiltonian}
\end{eqnarray}
with $\hat{{\vc r}} \equiv \frac{{\vc r}_1 - {\vc r}_2}{|{\vc r}_1 - {\vc r}_2|}$ being the unit vector of the relative coordinate ${\vc r}$.
The spin and isospin matrices are given by $\vc \sigma$ and $\tau$ ($\sigma_\pm \equiv \sigma_1 \pm \sigma_2$, $\tau_\pm \equiv \tau_1 \pm \tau_2$).
The subscripts 1 and 2 label the two interacting nucleons.
The radial function is given by
\begin{equation}
V_{CP}^{NN}(r)
= 
-\frac{m_\pi}{8\pi m_N} \frac{e^{-m_\pi r }}{r} \left( 1+ \frac{1}{m_\pi r} \right)
\ ,
\label{eq:cpvpotradial}
\end{equation}
with  $m_\pi = 138$ MeV and $m_N = 939$ MeV.
The shape of $V_{CP}^{NN}(r)$ is shown in Fig. \ref{fig:comparison_vcp_folded_alpha-triton}.
Since the CP-odd couplings are small, the nuclear EDM can linearly be expressed as
\begin{equation}
d_A 
=
\sum_{i=0,1,2} a_{A,\pi}^{(i)} \bar G_\pi^{(i)} 
,
\end{equation}
where $a_{A,\pi}^{(0,1,2)}$ depend only on the structure of the nucleus $A$.

The CP-odd $\alpha - t$ potential is calculated by applying the double folding to the bare CP-odd nuclear force.
It is given as follows:
\begin{eqnarray}
V_{CP}({\vc r}_t -{\vc r}_\alpha)
&=&
\bar G_\pi^{(1)}
\vc{\sigma_t} \cdot \frac{\vc{r}_t-\vc{r}_\alpha}{|\vc{r}_t-\vc{r}_\alpha|}
V_{CP}^{\alpha t }( |\vc{r}_t-\vc{r}_\alpha| ) 
,
\ \ \ 
\end{eqnarray}
where we explicitly wrote the coordinates of the triton and $\alpha$ cluster as $\vc{r}_t$ and $\vc{r}_\alpha$, respectively.
The radial function is
\begin{eqnarray}
V_{CP}^{\alpha t } 
( r ) 
&=&
\frac{1 }{\pi^2 m_N r^2}
\int_0^{\infty} d k \, 
\frac{ e^{- \frac{17}{48} k^2 b^2 }
}{k^2 + m_\pi^2}
\nonumber\\
&& \hspace{2em}\times
\bigl[
k^2 r \cos (kr)
-k \sin (k r )
\bigr]
,
\end{eqnarray}
where $b= 1.482$ fm.
The shape of $V_{CP}^{\alpha t}(r)$ is shown in Fig.~\ref{fig:comparison_vcp_folded_alpha-triton} (for the derivation, see Appendix \ref{sec:cpvtalpha}).
The isoscalar and isotensor CP-odd nuclear forces cancel after the folding.

We now briefly review the Gaussian expansion method \cite{hiyama}.
The wave function of the 2-body system ($^7$Li) is expressed as a superposition of the following Gaussian basis
\begin{eqnarray}
\Phi_{J=3/2} (^7{\rm Li} )
&=&
\sum_{nl}
C_{nl}^{^7{\rm Li},J}
[
\chi_{s,s_z} \otimes
\varphi_{nlm} ({\vc r}  )
]_{J=3/2}
,
\ \ \ \ 
\end{eqnarray}
where $\chi_{s,s_z} $ is the spin function of the triton cluster.
The Gaussian basis is defined as 
\begin{equation}
\varphi_{nlm} ({\vc r}  )
\equiv
N_{nl}
r^l e^{-\nu_n r^2}
Y_{lm} (\hat{r} )
.
\end{equation}
The radial function is expanded by gaussians with geometric series of the spreads $\lambda_n$.
The basis function of the 3-body system ($^{11}$B) is given by
\begin{eqnarray}
\Phi_{J=3/2} (^{11}{\rm B})
&=&
\sum_{n,N,l,L,\lambda }
A_{nNlL\lambda}^{^{11}{\rm B},J}
\Phi^{(23,1)}_{nNlL\lambda}
\nonumber\\
&&
+\sum_{n,N,l,L,\lambda } 
B_{nNlL\lambda}^{^{11}{\rm B},J}
\bigl[
\Phi^{(12,3)}_{nNlL\lambda}
+\Phi^{(31,2)}_{nNlL\lambda}
\bigr]
,
\nonumber\\
\end{eqnarray}
where the basis is expressed in terms of the Jacobi coordinates as $\Phi^{(ij,k)}_{nNlL\lambda} \equiv \bigl[ [\varphi_{nlm}({\vc r}_i -{\vc r}_j) \otimes \varphi_{NLM} ({\vc R}_k)]_\lambda \otimes \chi_{s,s_z} \bigr]_{J=3/2}$, with ${\vc R}_k \equiv {\vc r}_k- ( \sum_{i\neq k} m_i {\vc r}_i ) / ( \sum_{i\neq k} m_i ) $, $m_i$ being the mass of the $i$th cluster.

The Schr\"{o}dinger equations of the two- and three-body systems are solved by the variational principle,
\begin{eqnarray}
\delta\left[\langle \Phi_J \mid \mathcal{H}-E \mid \Phi_J \rangle\right]=0,
\label{eq:variational_principle}
\end{eqnarray}
where $E$ is the eigenenergy of $^7$Li ($^{11}$B) measured from the $\alpha+t$ ($\alpha+\alpha+t$) threshold.

\section{Results and discussion\label{sec:results}}

\subsection{A counting rule}

Let us first show the results of the calculation of the triton spin matrix elements.
For the $^7$Li, it is trivially unity,
\begin{equation}
\langle \, ^{7}{\rm Li} \, | \, \sigma_t \, |\,  ^{7}{\rm Li} \, \rangle
=
1.
\label{eq:7litritonspin}
\end{equation}
For the case of $^{11}$B, we numerically obtained
\begin{equation}
\langle \, ^{11}{\rm B} \, | \, \sigma_t \, |\,  ^{11}{\rm B} \, \rangle
=
0.76
.
\label{eq:11btritonspin}
\end{equation}
This matrix element is important since it is used to convolute the EDM of the $^3$H nucleus.
According to Ref. \cite{yamanakanuclearedm}, the EDM of $^3$H is 
\begin{eqnarray}
d_{^3{\rm H}}
&=&
0.88 d_p
-0.05 d_n
\nonumber\\
&&
-0.0059 \bar G_\pi^{(0)}
+0.0108 \bar G_\pi^{(1)}
-0.0170 \bar G_\pi^{(2)}
.
\label{eq:tritonedm}
\end{eqnarray}
From Eqs. (\ref{eq:7litritonspin}), (\ref{eq:11btritonspin}), and (\ref{eq:tritonedm}), we can calculate the contribution of the intrinsic nucleon EDM to the EDM of $^7$Li and $^{11}$B:
\begin{eqnarray}
d_{^7{\rm Li}}^{\rm (Nedm)}
&=&
0.9 d_p
-0.05 d_n
,
\\
d_{^{11}{\rm B}}^{\rm (Nedm)}
&=&
0.7 d_p
-0.04 d_n
.
\end{eqnarray}
We see that the coefficients (spin matrix elements) of the proton are close to unity for both $^7$Li and $^{11}$B.
This result is consistent with the simple shell model picture where the nucleus is composed of a core and a valence nucleon.
This is also consistent with the fact that the nuclear magnetic moments of $^7$Li ($\mu_{^{7}{\rm Li}} = +3.2564$) and $^{11}$B ($\mu_{^{11}{\rm B}} = +2.6886$) are close to the single proton one ($\mu_p = +2.7928$).
The positive values of $g-2$ of the above nuclei are important in experiment, since the measurement of the nuclear EDM using storage rings becomes easier \cite{storage1,storage2}.
The discrepancy of the spin matrix elements of the proton from unity is due to the configuration mixing, the superposition of several orbital angular momentum configurations.

Let us now show the results of the calculation of the polarization contribution in the $\alpha -t$ cluster model.
For the $^7$Li nucleus, we obtained
\begin{eqnarray}
d_{^{7}{\rm Li}}^{\rm (pol)} 
&=&
0.005 \, \bar G_\pi^{(1)} e\, {\rm fm}
+ d_{^3{\rm H}}^{\rm (pol)}
\nonumber\\
&=&
\Bigl(
-0.006 \, \bar G_\pi^{(0)}
+0.016 \, \bar G_\pi^{(1)} 
-0.017 \, \bar G_\pi^{(2)} 
\Bigr)e\, {\rm fm}
.
\nonumber\\
\label{eq:7liedmresult}
\end{eqnarray}
The first term was obtained by solving the Schr\"{o}dinger equation of the $\alpha -t$ system, and the second term is due to the EDM of the triton cluster.
Regarding the EDM of $^{11}$B, we found
\begin{eqnarray}
d_{^{11}{\rm B}}^{\rm (pol)} 
&=&
0.008 \, \bar G_\pi^{(1)} e\, {\rm fm}
+ 0.76 \, d_{^3{\rm H}}^{\rm (pol)}
\nonumber\\
&=&
\Bigl(
-0.004 \, \bar G_\pi^{(0)}
+0.016 \, \bar G_\pi^{(1)}
-0.013 \, \bar G_\pi^{(2)} 
\Bigr)e\, {\rm fm}
.
\nonumber\\
\label{eq:11bedmresult}
\end{eqnarray}
The coefficients are in accordance with the preliminary results reported in Refs. \cite{Yamanaka:2017tjl,Yamanaka:2018dwa}.
We improved their accuracy by taking 23 angular momentum channels with 4100 basis functions with which we found good convergence.
To estimate the theoretical uncertainty, we may use the deviation of the low-lying energy levels between our calculation and the experimental values (since the EDM is generated by the low-energy parity transitions).
Since they agree within 30\% \cite{yamada11b}, we may consider that the theoretical error bar is also at the same level.

\begin{figure*}[htb]
\begin{center}
\includegraphics[width=15cm]{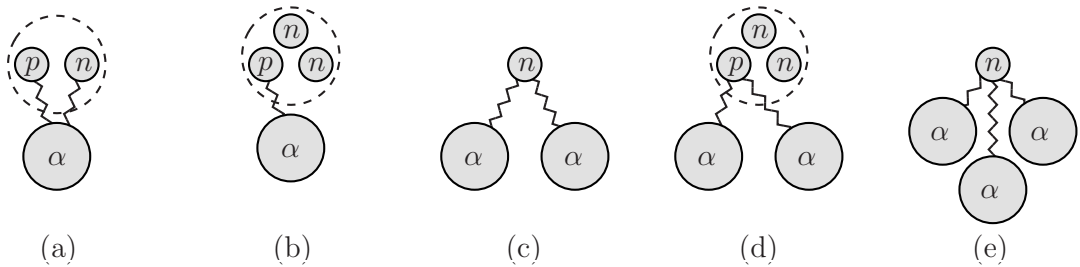}
\caption{\label{fig:counting_rule}
Schematic picture of the counting rule of the contribution to the EDM of (a) $^6$Li, (b) $^7$Li, (c) $^9$Be, (d) $^{11}$B, and (e) the 1/2$^+_1$ state of $^{13}$C.
The dashed circles indicate the deuteron (triton) cluster for the $^6$Li ($^7$Li or $^{11}$B) nucleus.
The CP-odd $\alpha - N$ polarization is denoted by zigzag lines.
The two neutrons in the triton cluster do not contribute to the CP-odd $\alpha - N$ polarization due to the dominant spin singlet configuration.
}
\end{center}
\end{figure*}

It is interesting to compare our results with the coefficients of other light nuclei known in the literature (see Table \ref{table:nuclearedm}).
By assuming that the EDM of light nuclei is generated by the constructive interference between that of the deuteron or $^3$H clusters and the polarization due to the CP-odd $\alpha-N$ polarization, we can solve a system of equations to determine the latter (see Fig. \ref{fig:counting_rule}):
\begin{eqnarray}
d_{^{6}{\rm Li}}
& = &
2\times (\alpha -N \, \mbox{polarization})
+ d_{^{2}{\rm H}}
,
\nonumber\\
d_{^{7}{\rm Li}}
& = &
1 \times (\alpha -N \, \mbox{polarization})
+d_{^{3}{\rm H}}
,
\nonumber\\
d_{^{9}{\rm Be}}
& = &
2 \times (\alpha -N \, \mbox{polarization})
,
\nonumber\\
d_{^{11}{\rm B}}
& = &
2 \times (\alpha -N \, \mbox{polarization})
+d_{^{3}{\rm H}}
,
\nonumber\\
d_{^{13}{\rm C}^*}
& = &
3 \times (\alpha -N \, \mbox{polarization})
.
\end{eqnarray}
After equating with the values of Table \ref{table:nuclearedm}, we obtain $(\alpha -N \, \mbox{polarization}) \sim (0.004 - 0.007)\, \bar G_\pi^{(1)}e$ fm.
This counting rule holds for light nuclei with a dominant contribution from maximally aligned angular momentum and spin.

From this counting rule, we can predict the EDMs of several stable $p$-and $sd$-shell nuclei.
For instance, we can infer
\begin{eqnarray}
d_{^{10}{\rm B}}
&\sim &
0.03\, \bar G_\pi^{(1)}e\, {\rm fm}
,
\\
d_{^{14}{\rm N}}
&\sim &
0.04\, \bar G_\pi^{(1)}e\, {\rm fm}
,
\\
d_{^{17}{\rm O}}
&\sim &
0.03\, \bar G_\pi^{(1)}e\, {\rm fm}
,
\\
d_{^{19}{\rm F}}
&\sim &
0.03\, \bar G_\pi^{(1)}e\, {\rm fm}
.
\end{eqnarray}
For the $^{15}$N nucleus, the configuration with the orbital angular momentum antiparallel with the spin of the valence nucleon is dominant, so that its EDM should be suppressed like for the case of the ground state of $^{13}$C \cite{c13edm}.

\begin{table}
\caption{
The linear coefficients of the CP-odd N-N coupling $a_\pi$ expressed in unit of $10^{-2} e$ fm.
}
\begin{ruledtabular}
\begin{tabular}{l|ccc}
  &$a_\pi^{(0)}$ & $a_\pi^{(1)}$ & $a_\pi^{(2)}$ \\ 
\hline
$^{2}$H \cite{liu,afnan,yamanakanuclearedm} & $-$ & $1.45 $ & $-$ \\
$^{3}$H \cite{bsaisou,yamanakanuclearedm} & $-0.59$ & 1.08 & $-1.70$ \\
$^{6}$Li \cite{yamanakanuclearedm} & $-$ & 2.2 & $-$  \\
$^{9}$Be \cite{yamanakanuclearedm} & $-$ & $1.4$ & $-$  \\
\vspace{0.2em}
$^{13}$C(1/2$^+_1$) \cite{c13edm} & $-$ & $2.4$ & $-$  \\
\hline
$^{7}$Li & $-0.6$ & 1.6 & $-1.7$  \\
$^{11}$B & $-0.4$ & 1.6 & $-1.3$  \\
\end{tabular}
\end{ruledtabular}
\label{table:nuclearedm}
\end{table}

\subsection{Chiral perturbation analysis}

Let us analyze the sensitivity to the new physics beyond the SM through the chiral perturbation theory.
The leading CP violating chiral Lagrangian contributing to the nuclear EDM is given by \cite{pospelovreview,Barton,pvcpvhamiltonian1,chiral3nucleon,bsaisou,yamanakanuclearedmreview,Gnech:2019dod}
\begin{eqnarray}
{\cal L}_{P\hspace{-.5em}/\, T\hspace{-.5em}/\, } 
&=& 
-\frac{i}{2} 
\bar d_N \bar N \sigma_{\mu \nu} \gamma_5 N F^{\mu \nu}
\nonumber\\
&&
+
\sum_{a=1}^3 
\Bigl[ \bar g_{\pi NN}^{(0)} \bar N \tau^a N \pi^a 
+ \bar g_{\pi NN}^{(1)} \bar N N \pi^0 
\Bigr]
\nonumber\\
&&
+m_N \Delta_{3\pi} \, \pi^z  \sum_{a=1}^3 \pi_a^2
\nonumber\\
&&
+
\bar C_1 \bar N N \partial_\mu (\bar N S^\mu N )
\nonumber\\
&&
+ \sum_{a=1}^3 \bar C_2 \bar N \tau_a N \cdot \partial_\mu (\bar N S^\mu \tau_a N )
, \ \ \ \ \ 
\label{eq:pcpvpinnint}
\end{eqnarray}
where $N^T= (p,n)$.
This chiral Lagrangian is defined with a cutoff $\Lambda \sim 1$ GeV.
We may additionally consider strange mesons and baryons with this cutoff, but their contribution should be negligible since the CP violation considered in this study is flavor diagonal.

We note that the nucleon EDMs $d_n$ and $d_p$ of Eq. (\ref{eq:intrinsisnucleonedm}) and the bare terms $\bar d_n$ and $\bar d_p$ given in Eq. (\ref{eq:pcpvpinnint}) are not identical.
In the leading order of chiral perturbation theory, $d_n$ and $d_p$ receive additional contribution from the CP-odd pion-nucleon interaction \cite{crewther}
\begin{eqnarray}
d_N
&=&
\bar d_N 
- \tau_z \frac{e g_A \bar g^{(0)}_{\pi NN} }{4 \pi^2 f_\pi} 
\ln \frac{\Lambda}{m_\pi}
,
\label{eq:d_1}
\end{eqnarray}
where $\tau_z = +1$ ($-1$) for the proton (neutron).

The CP-odd nuclear couplings of Eq. (\ref{eq:CPVhamiltonian}) can also be calculated from the chiral Lagrangian (\ref{eq:pcpvpinnint}).
In the leading order, they are given by
\begin{eqnarray}
\bar G_\pi^{(0)}
&=&
-\frac{g_A m_N}{f_\pi} \bar{g}_{\pi NN}^{(0)} , \\
\bar G_\pi^{(1)}
&=&
-\frac{g_A m_N }{f_\pi} 
\biggl[
\bar{g}_{\pi NN}^{(1)}
+\frac{15 g_A^2 m_\pi m_N}{32 \pi f_\pi^2} \Delta_{3\pi}
\biggr]
,
\label{eq:cpvcoupling}
\end{eqnarray}
where $g_A = 1.27$ and $f_\pi = 93$ MeV.
We note that the isovector coupling $\bar G_\pi^{(1)}$ generated by the three-pion interaction (term with $\Delta_{3\pi}$) has a momentum dependence, but here we neglect it since its effect is small \cite{bsaisou}.
The three-pion interaction can also generate a CP-odd three-nucleon force, but its effect vanishes in the $\alpha$ cluster model (the triton depends on it, but it is small \cite{bsaisou}).

In this work, we do not consider the isoscalar contact interactions [terms with $\bar C_1$ and $\bar C_2$ of Eq. (\ref{eq:pcpvpinnint})], since their effect has a large theoretical uncertainty \cite{bsaisou,bsaisou2}.
To control their effect, accurate nuclear wave functions at short distance are required.
We also have to note that $\bar C_1$ and $\bar C_2$ are the bare couplings defined at the scale $\Lambda \sim 1$ GeV with some renormalization scheme, and they will not stay the same in the cluster model, since the interactions vary under the change of the model space.
This change should be important for the short-range interaction, because the high-energy states with dissociated nucleons are integrated out in the cluster model.
We note that, although it has been neglected, the investigation of the isoscalar CP-odd contact interactions is important in the determination of the Weinberg operator \cite{weinbergop,Haisch:2019bml}, and it desirable to quantify this short-range physics in the future.

Combining Eqs. (\ref{eq:d_1}) and (\ref{eq:cpvcoupling}) with (\ref{eq:7litritonspin}), (\ref{eq:11btritonspin}), (\ref{eq:7liedmresult}), and (\ref{eq:11bedmresult}), we obtain
\begin{eqnarray}
d_{^{7}{\rm Li}} 
&=&
0.9 \bar d_p
-0.05 \bar d_n
\nonumber\\
&&
+\Bigl(
0.1 \, \bar g_{\pi NN}^{(0)}
-0.2 \, \bar g_{\pi NN}^{(1)}
-0.7 \Delta_{3\pi} 
\Bigr)e\, {\rm fm}
,
\\
d_{^{11}{\rm B}} 
&=&
0.7 \bar d_p
-0.04 \bar d_n
\nonumber\\
&&
+\Bigl(
0.1 \, \bar g_{\pi NN}^{(0)}
-0.2 \, \bar g_{\pi NN}^{(1)}
-0.8 \Delta_{3\pi}
\Bigr)e\, {\rm fm}
. \ \ \ \ \ 
\end{eqnarray}

\subsection{Impact on physics beyond the standard model}

We now analyze the prospects for the discovery of new physics beyond the SM.
We first inspect the contribution from the bare nucleon EDM ($\bar d_n$ and $\bar d_p$).
The quark level CP violation which is proper to it is the quark EDM.
The quark EDM is linearly related to the nucleon EDM through the nucleon tensor charge.
It has been evaluated in lattice QCD \cite{yaoki,jlqcd4,green,rqcdisovector,pndmeisovector,chiqcdisovector,etmsigmaterm,pndmetensor,egerer,Harris:2019bih}.
Its extractions from experimental data \cite{bacchetta,anselmino,courtoy,kang,radici,radici2} show tension with lattice QCD results, but this problem is expected to be resolved with future experiments \cite{yez,accardi,gaotensor}.
The effect of the quark EDM is suppressed by two factors.
First, the nucleon tensor charge is smaller than one, and does not show enhancement due to the dynamical effect of QCD in the nucleon \cite{yamanakasde1,pitschmann,hobbs}.
The other one is the suppression from the renormalization group evolution of the quark EDM \cite{tensorrenormalization,degrassi,yang}.
Typically, the evolution from the scale $\mu = 1$ TeV to $\mu = 1$ GeV suppresses the quark EDM by a factor of 0.8.
Although being suppressed, there are several models of new physics beyond the SM which only contribute through the quark EDM, such as the extended Higgs models with charged Higgs exchange \cite{kao,bowser-chao}, the split supersymmetry \cite{splitsusy1,splitsusy2,splitsusy3,splitsusy4} or some classes of R-parity violating supersymmetric models \cite{yamanakabook,rpv1,rpv3,rpvlinearprogramming}.
The sensitivity of the EDM of light nuclei on the bare nucleon EDM is therefore important in testing those models.

In genuine extended Higgs models or supersymmetric models, the quark chromo-EDM \cite{dekens,cedm} and the quark EDM are generated at the same time.
Since the quark chromo-EDM contributes to the CP-odd pion-nucleon interaction [second line of Eq. (\ref{eq:pcpvpinnint})], it is much more enhanced than the quark EDM contribution.

In the extended Higgs models, the leading contribution arises from the two-loop level Barr-Zee type diagram \cite{barr-zee,gunion,buras2hdmedm,brod,abe,inoue2hdmedm,jung2hdmedm,higgcisionedm,balazs,nakai,chenhiggs,bian,Panico,Brod:2018pli,Brod:2018lbf}.
For supersymmetric models, a large contribution is generated from the one-loop level diagram \cite{ellismssmedm,buchmuellermssmedm,polchinskimssmedm,delaguilamssmedm,nanopoulosmssmedm,duganmssmedm,kizukurimssmedm,fischlermssmedm,inuimssmedm,ibrahimmssmedm,pokorskimssmedm,pospelovreview}, together with the two-loop level diagrams \cite{chang,pilaftsis2loop,chang2loop,demir,feng2loop,mssmreloaded,liprofumo,mssmrainbow,pospelovreview}.
If the experiments reach the prospective sensitivity of the EDM of $O(10^{-29})\, e$ cm, it is possible to probe the scale of new physics beyond the SM up to $O(1-10)$ TeV, under the assumption of natural couplings and CP phases.
We note that the sensitivity to the supersymmetric CP phase of the $\mu$-term increases with large $\tan \beta$ \cite{pospelovreview,demir}.
In these models, many CP phases may appear at the same time, and it is important to examine the experimental data by taking into account the cancellation \cite{ibrahim,ramseyli,ellisgeometric,ellisgeometric2,rpvlinearprogramming,cancellationhiggs,cancellationhiggs2}.

In the left-right symmetric model, the isovector CP-odd pion-nucleon interaction [second term of the second line of Eq. (\ref{eq:pcpvpinnint})] is generated \cite{dekens}.
Since $\bar g_{\pi NN}^{(1)} \sim {\rm GeV}^2 / m_{W_R}^2$ \cite{atomicedmreview}, the $O(10^{-29})\, e$ cm sensitivity of the EDM of light nuclei studied in this paper can probe the right-handed weak boson of mass 1-10 PeV, within the natural $O(0.1)$ CP phase.

We also review the constraint on the QCD $\theta$-term.
The $\theta$-term contributes to the isoscalar CP-odd pion-nucleon interaction by $\bar g_{\pi NN}^{(0)} \sim 0.01 \bar \theta$ \cite{devries2}.
We then have $d_{^7{\rm Li}} , d_{^{11}{\rm B}} \sim 10^{-3} \bar \theta e$ fm, so that we may probe 
\begin{equation}
\bar \theta \sim O(10^{-13})
.
\end{equation}
under the prospective sensitivity of $d_{^7{\rm Li}} , d_{^{11}{\rm B}} \sim O(10^{-29})\, e$ cm. 
We note that the assumption of the Peccei-Quinn mechanism \cite{peccei} will unphysicalize $\bar \theta$, 
but instead it will be replaced by the coupling to the axion.
The sensitivity of the nuclear EDM of $^7$Li and $^{11}$B on the isoscalar CP violation $\bar g_{\pi NN}^{(0)}$ is therefore very important for the axion dark matter search.

Let us finally see the standard model CKM contribution.
Its effect arises in the second order of weak interaction, and the flavor change is crucial in the generation of the CP violation.
From the values of the CP-odd pion-nucleon couplings calculated in \cite{yamanakasmcpvnn}, the CKM contribution is predicted as
\begin{eqnarray}
d_{^7{\rm Li}} 
&\approx &
2 \times 10^{-31}e\, {\rm cm}
,
\\
d_{^{11}{\rm B}} 
&\approx &
2 \times 10^{-31}e\, {\rm cm}
,
\end{eqnarray}
where we neglected the contribution from the intrinsic nucleon EDM which is an order of magnitude smaller than the polarization contribution ($d_N \sim 10^{-32}e$ cm \cite{seng}).
These predictions are smaller than the sensitivity of the storage ring experiment, and proves the merit of the measurement of the EDM of $^7$Li or $^{11}$B.

\section{Summary}

In this paper we have calculated the EDM of $^7$Li and $^{11}$B in the cluster model with $^4$He and $^3$H clusters as degrees of freedom.
Our results show slightly larger isovector sensitivities than that of the deuteron.
According to our analysis, the nuclear polarization is generated by the constructive interference between the CP-odd $\alpha - N$ interaction and the EDM of $^3$H cluster which composes the system.

The $^7$Li and $^{11}$B nuclei have other interesting features than their sensitivity.
In the determination of the new physics beyond the SM, they may be interesting from the point-of-view of the nonorthogonality, since they receive CP-odd isoscalar contributions from the intrinsic nucleon EDM as well as from the polarization of the triton cluster, as opposed to the deuteron or $^6$Li which is insensitive to it.
This point is also potentially important in the detection of the axion dark matter which replaces the effect of the $\theta$-term.
This is indeed a very attractive approach since the interaction with the axion may induce an observable oscillating nuclear EDM which avoids Schiff's screening.
The potentiality has definitely to be studied in the future.
We also note that $^7$Li and $^{11}$B have both positive $g-2$, which may be important in experimental measurements using storage rings.

From our study, a new phenomenon due to the clustering was also suggested.
Our results, together with the previous predictions of the EDM of $^6$Li, $^9$Be, and the $1/2_1^+$ state of $^{13}$C, suggest a counting rule consisting of adding the EDM of the constituent cluster (deuteron or $^3$H) and the polarization due to the CP-odd $\alpha - N$ interaction which takes an approximate value $a^{(1)}_{\rm pol} \sim (0.004 -0.007 ) \bar G_\pi^{(1)} e$ fm.
Using this counting rule, it is possible to predict the EDM of other $p$-shell nuclei.
One interesting candidate is the $^{14}$N nucleus, for which the isovector sensitivity may reach $d_{^{14}{\rm N}} \sim 0.04 \bar G_\pi^{(1)} e$ fm.
We expect these features to be unveiled in future works.

\begin{acknowledgments}
This work is supported by the Grant-in-Aid for Scientific Research [No.18K03629 and No.18K03658] from the Japan Society For the Promotion of Science.
NY was supported by the JSPS Postdoctoral Fellowships for Research Abroad.
\end{acknowledgments}

\appendix

\section{\label{sec:cpvtalpha}Derivation of the CP-odd $\alpha - t$ interaction in OCM}

We derive the contribution of the isovector CP-odd nuclear force [see Eq. (\ref{eq:CPVhamiltonian})] to the CP-odd $\alpha -t$ interaction by folding.
The direct potential between the triton and alpha clusters can be written as follows:
\begin{eqnarray}
&& \langle
\psi (t) \psi (\alpha ) \, |
\sum_{i \in t , j \in \alpha } V_{CP}^{NN} ( |{\vc r}_i - {\vc r}_j |) \times
\nonumber\\
&& \hspace{3em}
(\tau_i^z \vc{\sigma}_i -\tau_j^z \vc{\sigma}_j) \cdot \frac{\vc{r}_i-\vc{r}_j}{|\vc{r}_i-\vc{r}_j|}
|\, 
\psi (t) \psi (\alpha)
\rangle
\nonumber\\
&=&
\langle
\phi (t) \phi (\alpha) \, |
\sum_{i \in t , j \in \alpha } 
V_{CP}^{NN}( |{\vc r}_i - {\vc r}_j |)\times
\nonumber\\
&& \hspace{3em}
\tau_i^z \vc{\sigma}_i \cdot \frac{\vc{r}_i-\vc{r}_j}{|\vc{r}_i-\vc{r}_j|}
|\, 
\phi (t) \phi (\alpha)
\rangle
\nonumber\\
&=&
\langle
\phi (t) \phi (\alpha) \, |
\nonumber\\
&& \hspace{0em}\times
\sum_{i \in t , j \in \alpha }  V_{CP}^{NN} (\, | {\vc r}_i - {\vc X}_t - [ {\vc r}_j -{\vc X}_{\alpha} ] + {\vc r} |\, )
\nonumber\\
&& \hspace{0em}\times
\tau_i^z  \vc{\sigma}_i \cdot \frac{ {\vc r}_i - {\vc X}_t - [ {\vc r}_j -{\vc X}_{\alpha} ] + {\vc r}}{| {\vc r}_i - {\vc X}_t - [ {\vc r}_j -{\vc X}_{\alpha} ] + {\vc r} |}
\,
|\, \phi (t) \phi (\alpha )
\rangle \nonumber \\
&\equiv & W,
\label{eq:A1}
\end{eqnarray}
where $\vc{X}_t$ and $\vc{X}_\alpha$ are the center-of-mass coordinates of the triton and $\alpha$ clusters, respectively, and $\vc{r}= {\vc X}_{t} - {\vc X}_{\alpha}$.
In the first equality, we use the fact that the spin and isospin matrices acting on the nucleons of the $\alpha$ cluster (term with $\tau_j^z \vc{\sigma}_j$, $j \in \alpha$) cancel due to the spin and isospin saturation property of the $\alpha$ cluster.
The isoscalar and the isotensor CP-odd nuclear forces vanish for the case of the $\alpha -t $ system, since the isospin Pauli matrices operate to both interacting nucleons.

It is then convenient to use the following transformation of the CP-odd nuclear force (\ref{eq:cpvpotradial}),
\begin{eqnarray}
V_{CP}^{NN} ( R )
\hat{\vc R}
&=&
- \frac{m_\pi}{8 \pi m_N } \frac{e^{-m_\pi R }}{R} \left( 1+ \frac{1}{m_\pi r} \right)
\hat{\vc R}
\nonumber\\
&=&
\frac{1}{2 m_N}
\vc{\nabla} \frac{e^{-m_\pi |{\vc R}|}}{4\pi |{\vc R}|}
\nonumber\\
&=&
\frac{1}{2 m_N}
\vc{\nabla} 
\int \frac{d {\vc k}}{(2\pi )^3}
\frac{ e^{i {\vc k} \cdot {\vc R} }}{|{\vc k}|^2 +m_\pi^2}
\nonumber\\
&=&
\frac{1}{2 m_N}
\int \frac{d {\vc k}}{(2\pi )^3}
\frac{i{\vc k}\, e^{i {\vc k} \cdot {\vc R} }}{|{\vc k}|^2 +m_\pi^2}
\nonumber\\
&=&
\int \frac{d {\vc k}}{(2\pi )^3}
{\cal F}(k) \hat{\vc{k}} e^{i {\vc k} \cdot {\vc R} },
\label{eq:pionexchangeFT}
\end{eqnarray}
where ${\cal F}(k)\equiv ik/[2m_N (k^2+m_\pi^2)]$. Here we clearly see the structure induced by the one-pion exchange. This transformation makes clear the relation of the CP-odd nuclear couplings with the chiral perturbation formula of Eq.~(\ref{eq:cpvcoupling}).
By substituting the above formula into Eq.~(\ref{eq:A1}), the direct potential $W$ can be expressed as follows:
\begin{eqnarray}
W
&=&
\int \frac{d {\vc k}}{(2\pi)^3} \,
\sum_{i \in t , j \in \alpha } \langle \phi (t) \phi (\alpha) \, |
\nonumber \\
&& \times
{\cal F}(k) \tau_i^z \vc{\sigma}_i \cdot \vc{\hat k}
e^{i {\vc k} \cdot ({\vc r}_i - {\vc X}_t - [ {\vc r}_j -{\vc X}_{\alpha} ] + {\vc r} ) } 
|\, \phi (t) \phi (\alpha )
\rangle
\nonumber \\
&=&
\int \frac{d {\vc k}}{(2\pi)^3} e^{i\vc{k}\cdot \vc{r}} {\cal F}(k) \rho_t(\vc{k}) \rho_\alpha (\vc{k})^* ,
\label{eq:A3}
\end{eqnarray}
where
\begin{eqnarray}
\rho_t(\vc{k})
&=&  \langle \phi (t) \, | \sum_{i \in t} \tau_i^z \vc{\sigma}_i \cdot \vc{\hat k} 
e^{i {\vc k} \cdot ({\vc r}_i - {\vc X}_t) } |\, \phi (t) \rangle , 
\label{eq:A4_1} 
\\
\rho_\alpha (\vc{k}) 
&=& 
\langle  \phi (\alpha) \, |  \sum_{j \in \alpha}
e^{i {\vc k} \cdot ({\vc r}_j -{\vc X}_{\alpha}) } 
\, |\, \phi (\alpha ) \rangle .
\label{eq:A4_2}
\end{eqnarray}
In general, the Fourier transformation of the one-body density distribution of $0s$-shell cluster $C$ with mass number $A\leq 4$ can easily be calculated as follows:
\begin{eqnarray}
\rho_{C} ({\vc k}) 
&= &
\langle
\phi (C)  \, |
\sum_{i=1}^A 
e^{i {\vc k} \cdot ( {\vc r}_i -{\vc X}_{C})  } 
\,
|\, \phi (C) 
\rangle
\nonumber\\
&=&
\langle
\phi_G ({\vc X}_{C}) \phi (C)  \, |
e^{i {\vc k} \cdot {\vc X}_{C}  } 
\sum_{i=1}^A
e^{i {\vc k} \cdot ( {\vc r}_i -{\vc X}_{C})  } 
\,
\nonumber\\
&& \hspace{8em}\times
|\, \phi_G ({\vc X}_{C}) \phi (C) 
\rangle
\nonumber\\
&& \hspace{2em}/ 
\langle
\phi_G ({\vc X}_{C} )  \, |
e^{i {\vc k} \cdot {\vc X}_{C}  } 
\,
|\, \phi_G ({\vc X}_{C}) 
\rangle
\nonumber\\
&=&
e^{\frac{b^2 k^2}{4 A }}
\langle
\phi_G ({\vc X}_{C}) \phi (C)  \, |
\sum_{i=1}^A
e^{i {\vc k} \cdot {\vc r}_i  } 
\,
|\, \phi_G ({\vc X}_{C}) \phi (C) 
\rangle
\nonumber\\
&=&
e^{\frac{b^2 k^2}{4 A}}
\langle
\Psi (C)  \, |
\sum_{i=1}^A
e^{i {\vc k} \cdot {\vc r}_i  } 
\,
|\, \Psi (C) 
\rangle
\nonumber\\
&=&
Ae^{\frac{b^2 k^2}{4 A}}
\langle
\varphi_{0s} ({\vc r})  \, |
e^{i {\vc k} \cdot {\vc r}  } 
\,
|\, \varphi_{0s} ({\vc r}) 
\rangle
\nonumber \\
&=&
A\exp\Big(-\frac{b^2 k^2}{4}\frac{A-1}{A} \Big),
\end{eqnarray}
where $\Psi (C) $ is the Slater determinant composed of $A$ single-particle states, and $\phi_G (\vc{X_C})$ and $\varphi_{0s}(\vc{r})$ are the center-of-mass wave function and the $0s$ single-particle wave function, respectively, expressed below.
\begin{eqnarray}
\phi_G (\vc{X}_C) &=& \Big(\frac{\pi b^2}{A}\Big)^{-3/4} \exp \Big(-\frac{A}{2b^2} \vc{X}_C^2 \Big), 
\nonumber \\
\varphi_{0s}(\vc{r}) &=& (\pi b^2)^{-3/4} \exp \Big(-\frac{\vc{r}^2}{2b^2}\Big) .
\end{eqnarray}
The one-body distribution of the $\alpha$ cluster, 
$\rho_\alpha (\vc{k})$, with $A=4$, is then simply given by 
\begin{equation}
\rho_{\alpha} (\vc{k}) =
4\exp \Big( -\frac{3b^2}{16} k^2 \Big).
\label{eq:rho_al}
\end{equation}
That of the triton, $\rho_t(\vc{k})$ in Eq.~(\ref{eq:A4_1}), can also be calculated in a similar way, by noticing that the contribution from the two neutrons in the triton cancels, due to the spin shell closure property, resulting in
\begin{eqnarray}
\rho_{t} ({\vc k}) 
&=& 
\exp \Big(-\frac{b^2 k^2}{6}\Big) \vc{\sigma}_t \cdot \hat{\vc{k}},
\label{eq:rho_tri}
\end{eqnarray}
where the contribution from the spin of the proton (the third nucleon) only remains, with $\vc{\sigma}_t \equiv \langle \vc{\sigma}_3 \rangle$.

Inserting Eqs.~(\ref{eq:rho_al}) and (\ref{eq:rho_tri}) into Eq.~(\ref{eq:A3}), the folding potential between the $\alpha$ and triton clusters is obtained as
\begin{eqnarray}
W
&=&
\frac{4i}{2 m_N(2 \pi)^3}
\int d^3 {\vc k} \, e^{i {\vc k} \cdot {\vc r}} \frac{\vc{\sigma}_t \cdot \vc{k}}{k^2 + m_\pi^2}
e^{-\frac{17}{48}b^2k^2}
\nonumber\\
&=&
\frac{i \vc{\sigma}_t \cdot \hat{{\vc r}}}{2 \pi^2 m_N}
\int_0^\infty k^2 d k \, 
\frac{ke^{-\frac{17}{48}b^2k^2}}{k^2 + m_\pi^2} 
\int_{-1}^1 d (\cos \theta) \cos \theta \,
e^{i kr \cos \theta}
\nonumber\\
&=&
-\frac{i \vc{\sigma}_t \cdot \hat{{\vc r}}}{2 \pi^2 m_N r^2}
\int_{-\infty}^{\infty} k d k \, 
\frac{e^{-\frac{17}{48}b^2k^2}}{k^2 + m_\pi^2}
e^{-i kr } (1+ikr)
\nonumber\\
&=&
\frac{\vc{\sigma}_t \cdot \hat{{\vc r}} }{\pi^2 m_N r^2}
\int_0^{\infty} d k \, 
\frac{ e^{- \frac{17}{48}k^2 b^2 }
}{k^2 + m_\pi^2}
\Bigl[
k^2 r \cos (kr)
-k \sin (k r )
\Bigr]
\nonumber \\
&\equiv& V_{CP}^{\alpha t} ( r ) \vc{\sigma}_t \cdot \hat{\vc r}
.
\end{eqnarray}

\end{document}